\documentclass[]{elsarticle}
\usepackage{graphicx,natbib,amssymb}

\begin{document}

\begin{frontmatter}

   \title{Abundance analysis of a sample of evolved stars in the outskirts of $\omega$ Centauri\thanks{Based on data collected at ESO VLT under program 267.D-5695(A)}}

\author[lbl1]{S. Villanova}
\author[lbl2]{G. Carraro}
\ead{gcarraro@eso.org}
\author[lbl3]{R. Scarpa}
\author[lbl2]{G. Marconi}

\address[lbl1]{Universidad de Concepci\'on, Departamento de Astronomia, Casilla
              160-C, Concepci\'on, Chile}

\address[lbl2]{ESO, Alonso de Cordova 3107, Vitacura, Santiago de Chile, Chile}

\address[lbl3]{Instituto de Astrofisica de Canarias, La Laguna, Tenerife,
               Spain}

\begin{abstract}
   The globular cluster $\omega$ Centauri (NGC~5139) is a puzzling stellar system
   harboring several distinct stellar populations whose origin still represents a 
   unique astrophysical challenge. Current scenarios range from primordial chemical inhomogeneities 
   in the mother cloud to merging of different sub-units and/or subsequent generations
   of enriched stars - with a variety of different pollution sources- within the same potential well. 
   In this paper we study the chemical abundance pattern in the outskirts of $\omega$~ Centauri, 
   half-way to the tidal radius (covering the range of 20-30 arcmin from the cluster center), 
   and compare it with chemical trends in the inner cluster regions, in an attempt to
   explore whether the same population mix and chemical compositions
   trends routinely found in the more central regions 
   is also present in the cluster periphery.
   We extract abundances of many elements from FLAMES/UVES spectra of 48 RGB stars
   using the equivalent width method and
   then analyze the metallicity distribution function and abundance ratios of the observed stars.
   We find, within the uncertainties of small number statistics and slightly different evolutionary phases, 
   that the population mix in the outer regions cannot be distinguished
   from the more central regions, although it is clear that more data are necessary to
   obtain a firmer description of the situation. 
   From the abundance analysis, we did not find obvious radial gradients in any of
   the measured elements.
\end{abstract}


\end{frontmatter}

\section{Introduction} 

Multiple stellar populations are routinely 
found in old Galactic and intermediate-age Magellanic Clouds star clusters
(Piotto 2008 and references therein). 
Whether they are a signature of the cluster formation process or a result of the star formation
history and related stellar
evolution effects, is still matter of lively discussion (Renzini 2008, Bekki et al. 2008, Decressin et al. 2007).
The prototype of globular hosting multiple populations has for long time been 
$\omega$~ Cen (Villanova et al. 2007), although the current understanding is that it is possibly the remnant
of a dwarf galaxy (Carraro \& Lia 2000, Tsuchiya et al. 2004, Romano et al. 2007).\\
Most chemical studies of the stellar population in $\omega$~ Cen are
restricted within 20 arcmin of the cluster radius center (see Norris \& Da Costa 1995, Villanova et al. 2007),
 where, probably, the diverse stellar components are better mixed and
less subjected to external perturbations, like the Galactic tidal stress, than the outer
regions. Assessing whether there are population inhomogeneities in $\omega$ Cen or gradients
in metal abundance is a crucial step to progress in our understanding of this fascinating stellar
system.\\
In Scarpa et al. (2003, 2007) we presented the results of a spectroscopic campaign to
study the stellar radial velocity dispersion profile at $\sim$ 25 arcmin, half way to  
the tidal radius ($\sim$ 57 arcmin,  Harris 1996), in an attempt to find a new way to verify the 
MOND (Modified Newtonian Dynamics, Milgrom 1983) theory of gravitation.\\
In this paper we make use of a subsample of those spectra (the ones taken for
RGB stars) and extract estimates of metal abundances for some of the most
interesting elements.
The aim is to study the chemical trends of the stellar populations in the cluster periphery,
to try to learn whether the cluster outskirts contain, both qualitatively
and quantitatively, the same population mix and to infer from this additional information
on the cluster formation and evolution.\\
The layout of the paper is as follows. In Sect.~2 we describe observations and data reduction,
while Sect.~3 is dedicated to the derivation of metal abundances.
These latter are then discussed in detail in Sect.~4. Sect.~5 is devoted to the comparison
of the metal abundance trends in the inner and outer regions of $\omega$~ Cen, and, finally,
Sect.~6 summarizes the findings of this study.

\begin{figure}
\centering
\includegraphics[width=\columnwidth]{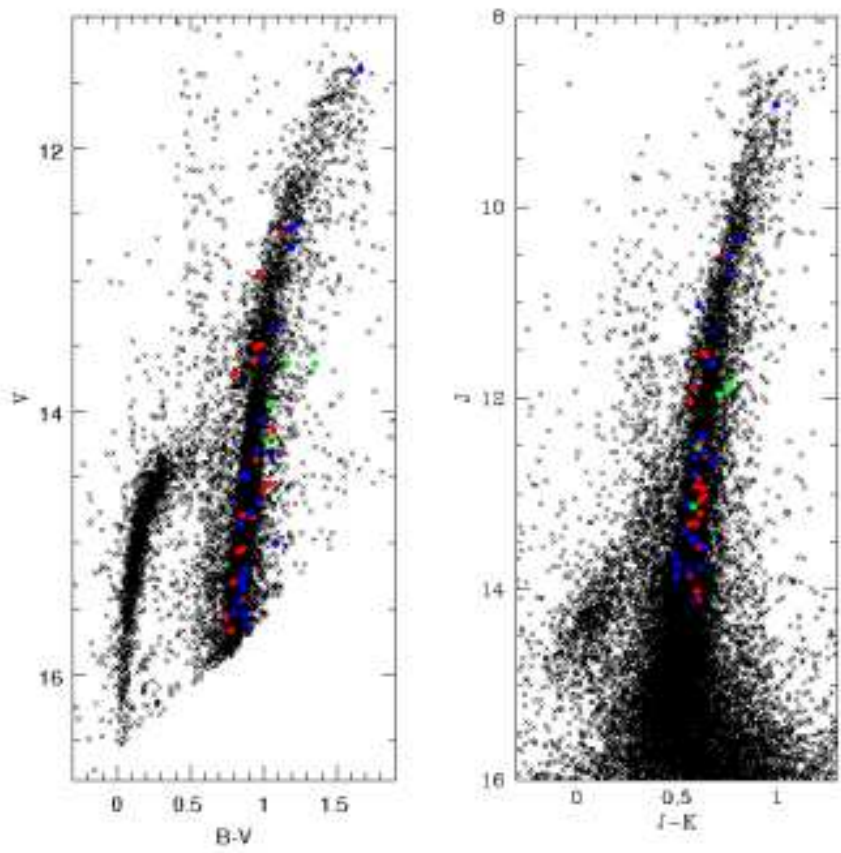}
\caption{The CMD of $\omega$~ Cen in the optical (left panel) and infrared.
Solid symbols of different colors indicate stars belonging to the MPP (red),
IMP (blue) and MRP(green). See text for more details.}
\label{f1}
\end{figure}

\begin{figure}
\centering
\includegraphics[width=\columnwidth]{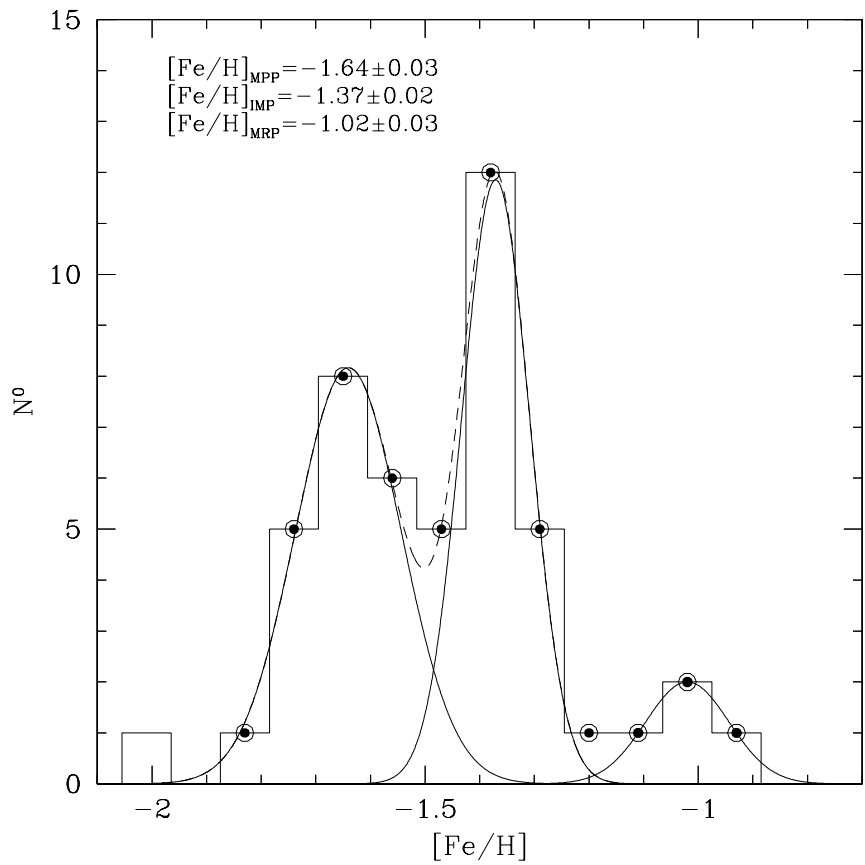}
\caption{Distribution of iron abundance for the program stars. A bimodal
Gaussian fit is used to derive the mean iron abundance of the MPP and IMP. 
Mean iron abundances of the three peaks are indicated. See text for more details.}
\label{f2}
\end{figure}

\begin{table} [!hbp]
\caption{Measured Solar abundances ($\rm
  {log\epsilon(X)=log(N_{X}/N_{H})+12)}$.} 
\label{t1}
\centering
\begin{tabular}{lc}
\hline
\hline
Element &  log$\epsilon$(X)\\
\hline
NaI & 6.37 \\
MgI & 7.54 \\
SiI & 7.61 \\
CaI & 6.39 \\
TiI & 4.94 \\
TiI & 4.96 \\
CrI & 5.63 \\
FeI & 7.50 \\
NiI & 6.28 \\
ZnI & 4.61 \\
YII & 2.25 \\
BaI & 2.31 \\
\hline
\end{tabular}
\end{table}
\noindent


\section{Observations and Data reduction}
Our data-set consists of UVES spectra collected in
August 2001, for a project devoted to 
measuring radial velocities and establishing membership in the outskirts of the 
cluster. Data were obtained with UVES/VLT@UT2
(Pasquini et al.\ 2002) with a typical seeing of 0.8-1.2 arcsec. 
We observed isolated stars from the lower red giant branch (RGB) up to the tip
of the RGB of $\omega$ Cen, in the magnitude
range $11.5<{\rm V}<16.0$.

We used the UVES spectrograph in the RED 580 setting. The spectra have a spectral coverage of
$\sim$2000 \AA \ with the central wavelength at 5800 \AA. The typical
signal to noise ratio is ${\rm S/N\sim 20-30}$.
For additional details, the reader is referred to Scarpa et al. (2003).

Data were reduced using UVES pipelines (Ballester et al.\ 2000),
including bias subtraction, flat-field correction, wavelength
calibration, sky subtraction and spectral rectification. Stars
were selected from photographic BV observations (van Leeuwen et al. 2000)
coupled with infrared JHK 2MASS photometry (Skrutskie
et al.\ 2006). Targets are located at a radial distance between 20 and 30
arcmin. The whole sample of stars contain both RGB and horizontal branch (HB) stars.
In this paper we focus our attention only on RGB objects, for the sake of comparison
with previous studies.

\subsection{Radial velocities and membership}
In the present work, radial velocities were used as the membership criterion
since the cluster stars all have similar motions with respect to the observer.
The radial velocities of the
stars were measured using the IRAF FXCOR task, which cross-correlates
the object spectrum with a template.  As a template, we used a
synthetic spectrum obtained through the spectral synthesis code
SPECTRUM (see {\sf http://www.phys.appstate.edu/spectrum/spectrum.html} for
more details), using a Kurucz model atmosphere with roughly the mean
atmospheric parameters of our stars $\rm {T_{eff}=4900}$ K, $\rm {log(g)=2.0}$,
$\rm {v_{t}=1.3}$ km/s, $\rm {[Fe/H]=-1.40}$. Each radial velocity was
corrected to the heliocentric system. We calculated a first approximation
mean velocity and the r.m.s ($\sigma$) of the velocity distribution.
Stars showing $\rm {v_{r}}$ out of more than
$3\sigma$ from the mean value were considered probable field objects and
rejected, leaving us with 48 UVES spectra of probable members, whose position
in the CMD are shown in Fig.~\ref{f1}. Radial velocities
for member stars are reported in Tab.~\ref{t2}

\section{Abundance analysis}
\subsection{Continuum determination}

The chemical abundances for all elements
were obtained from the equivalent widths (EWs) of the spectral lines (see next
Section for the description of the line-list we used).  
An accurate measurement of EWs first requires a good determination
of the continuum level. Our relatively metal-poor stars
allowed us to proceed in the following way. First, 
for each line, we selected a region of 20 \AA \ centered on the line itself
(this value is a good compromise between having enough points, i. e. a good statistic, and 
avoiding a too large region where the spectrum might not be flat).
Then we built the histogram of the distribution of the flux where the peak is a
rough estimation of the continuum. We refined this determination by fitting a
parabolic curve to the peak and using the vertex as our continuum estimation. 
Finally, we revised the continuum determination by eye and corrected by hand
if an obvious discrepancy with the spectrum was  found.
Then, using the continuum value previously obtained, we fit a Gaussian curve
to each spectral line and obtained the EW from integration.
We rejected lines if they were affected by bad continuum determinations, by non-Gaussian
shape, if their central wavelength did not agree with that expected from 
our line-list, or if the lines were too broad or too narrow with respect to the
mean FWHM.
We verified that the Gaussian shape is a good approximation for our spectral
lines, so no Lorenzian correction has been applied.

\begin{table*}
\caption{Stellar parameters. Coordinates are for J2000.0 equinox}
\label{t2}
\centering
\begin{tabular}{cccccccccccc}
\hline
\hline
ID & $\alpha$ & $\delta$ & B & V & J$_{\rm 2MASS}$ & H$_{\rm 2MASS}$ & K$_{\rm 2MASS}$ & T$_{\rm eff}$& log(g) & v$_{\rm t}$ & RV$_{\rm H}$\\
\hline
& deg & deg. &  & & & & & $^{0}K$ & & km/sec & km/sec \\
\hline                      
00006 & 201.27504 & -47.15599 & 16.327 & 15.531 & 13.865 & 13.386 & 13.364 & 5277 & 2.75 & 1.23 & 222.99\\
08004 & 201.07113 & -47.22082 & 15.393 & 14.508 & 12.687 & 12.110 & 12.007 & 4900 & 2.17 & 1.38 & 241.70\\
10006 & 201.16314 & -47.23357 & 14.510 & 13.710 & 11.887 & 11.413 & 11.300 & 5080 & 1.93 & 1.44 & 237.18\\
10009 & 201.24457 & -47.23406 & 13.807 & 12.573 & 10.331 &  9.664 &  9.520 & 4432 & 1.14 & 1.64 & 227.57\\
10010 & 201.33458 & -47.23334 & 14.982 & 13.941 & 11.963 & 11.394 & 11.249 & 4758 & 1.88 & 1.45 & 220.18\\
13006 & 201.13373 & -47.25880 & 16.442 & 15.665 & 14.112 & 13.615 & 13.504 & 5251 & 2.79 & 1.22 & 231.78\\
14002 & 201.16243 & -47.26471 & 15.696 & 14.853 & 13.110 & 12.634 & 12.552 & 5151 & 2.42 & 1.31 & 224.00\\
22007 & 201.08521 & -47.32639 & 14.799 & 13.635 & 11.843 & 11.221 & 11.077 & 4750 & 1.75 & 1.49 & 227.25\\
25004 & 201.18696 & -47.34607 & 15.048 & 14.064 & 12.393 & 11.852 & 11.762 & 5034 & 2.06 & 1.41 & 230.68\\
27008 & 201.16507 & -47.36326 & 15.242 & 14.220 & 12.519 & 12.046 & 11.911 & 5095 & 2.15 & 1.38 & 237.50\\
28009 & 201.13729 & -47.36499 & 15.687 & 14.779 & 13.133 & 12.664 & 12.549 & 5186 & 2.41 & 1.32 & 236.00\\
33006 & 201.12822 & -47.40730 & 13.062 & 11.403 &  8.924 &  8.064 &  7.929 & 4051 & 0.39 & 1.83 & 226.34\\  
34008 & 201.19496 & -47.41343 & 13.803 & 12.629 & 10.510 &  9.897 &  9.749 & 4570 & 1.25 & 1.61 & 232.16\\
38006 & 201.11643 & -47.44354 & 16.289 & 15.436 & 13.822 & 13.304 & 13.263 & 5202 & 2.68 & 1.25 & 222.58\\
39013 & 201.16078 & -47.45089 & 13.950 & 12.755 & 10.690 & 10.097 &  9.935 & 4610 & 1.32 & 1.59 & 231.47\\
42012 & 201.17440 & -47.47487 & 14.468 & 13.379 & 11.299 & 10.705 & 10.613 & 4673 & 1.61 & 1.52 & 232.58\\
43002 & 201.14213 & -47.47916 & 15.313 & 14.365 & 12.597 & 12.113 & 11.956 & 5021 & 2.17 & 1.38 & 229.17\\
45011 & 201.10941 & -47.49389 & 16.208 & 15.346 & 13.630 & 13.146 & 13.146 & 5229 & 2.66 & 1.25 & 224.37\\
45014 & 201.15625 & -47.50013 & 15.894 & 15.066 & 13.316 & 12.803 & 12.720 & 5073 & 2.47 & 1.30 & 249.24\\
46003 & 201.12943 & -47.50252 & 15.640 & 14.788 & 13.073 & 12.578 & 12.455 & 5091 & 2.37 & 1.33 & 242.00\\
48009 & 201.12036 & -47.51844 & 16.504 & 15.616 & 14.125 & 13.602 & 13.537 & 5279 & 2.79 & 1.22 & 222.94\\
49008 & 201.16235 & -47.52717 & 15.657 & 14.687 & 12.799 & 12.256 & 12.210 & 4925 & 2.26 & 1.36 & 238.57\\
51005 & 201.09190 & -47.53945 & 16.140 & 15.292 & 13.551 & 13.005 & 12.913 & 5028 & 2.55 & 1.28 & 221.27\\
57006 & 201.18559 & -47.58523 & 15.906 & 15.046 & 13.320 & 12.797 & 12.757 & 5096 & 2.48 & 1.30 & 234.33\\
61009 & 201.16032 & -47.61620 & 14.488 & 13.496 & 11.533 & 10.947 & 10.890 & 4784 & 1.71 & 1.50 & 239.65\\
76015 & 201.33839 & -47.73435 & 15.602 & 14.604 & 12.839 & 12.355 & 12.231 & 5026 & 2.27 & 1.35 & 241.73\\
77010 & 201.23548 & -47.74124 & 14.992 & 13.641 & 11.886 & 11.269 & 11.133 & 4746 & 1.75 & 1.49 & 238.49\\
78008 & 201.21908 & -47.74676 & 16.088 & 15.001 & 13.484 & 12.990 & 12.909 & 5221 & 2.52 & 1.29 & 223.14\\
80017 & 201.40179 & -47.75878 & 15.250 & 14.294 & 12.481 & 11.989 & 11.896 & 5026 & 2.15 & 1.38 & 231.59\\
82012 & 201.44193 & -47.77921 & 16.094 & 15.298 & 13.558 & 13.059 & 12.947 & 5099 & 2.58 & 1.27 & 232.28\\
85007 & 201.19307 & -47.80062 &   -    &   -    & 14.020 & 13.489 & 13.419 & 4983 & 2.20 & 1.37 & 250.48\\
85014 & 201.37723 & -47.80134 & 15.400 & 14.347 & 12.560 & 11.982 & 11.923 & 4899 & 2.11 & 1.39 & 236.78\\
85019 & 201.53965 & -47.80194 & 15.727 & 14.803 & 12.939 & 12.428 & 12.308 & 4938 & 2.31 & 1.34 & 243.81\\
86007 & 201.22490 & -47.80442 &   -    &   -    & 13.024 & 12.487 & 12.387 & 4914 & 1.88 & 1.45 & 238.70\\
86010 & 201.31217 & -47.80789 & 15.594 & 14.557 & 12.926 & 12.437 & 12.329 & 5115 & 2.29 & 1.35 & 238.05\\
86017 & 201.56208 & -47.80760 & 16.289 & 15.452 & 13.737 & 13.319 & 13.232 & 5290 & 2.73 & 1.24 & 231.74\\
87009 & 201.61710 & -47.81630 & 16.081 & 15.199 & 13.392 & 12.885 & 12.850 & 5082 & 2.54 & 1.29 & 247.67\\
88023 & 201.58521 & -47.82029 & 16.415 & 15.542 & 13.774 & 13.268 & 13.154 & 5050 & 2.66 & 1.25 & 232.87\\
89009 & 201.57067 & -47.83291 & 13.776 & 12.650 & 10.497 &  9.890 &  9.753 & 4568 & 1.25 & 1.61 & 242.07\\
89014 & 201.66544 & -47.83110 & 14.611 & 13.607 & 11.639 & 11.055 & 10.967 & 4774 & 1.75 & 1.49 & 231.57\\
90008 & 201.22516 & -47.83980 &   -    &   -    & 13.209 & 12.703 & 12.591 & 5010 & 1.95 & 1.43 & 240.42\\
90019 & 201.62529 & -47.83825 & 14.462 & 13.509 & 11.537 & 11.018 & 10.911 & 4860 & 1.75 & 1.48 & 232.73\\
90020 & 201.64363 & -47.83814 & 16.305 & 15.563 & 13.858 & 13.395 & 13.292 & 5219 & 2.74 & 1.23 & 240.39\\
93016 & 201.65058 & -47.86211 & 15.342 & 14.479 & 12.620 & 12.107 & 12.031 & 5015 & 2.22 & 1.37 & 230.82\\
94011 & 201.30980 & -47.86480 & 15.217 & 14.151 & 12.462 & 11.911 & 11.842 & 4989 & 2.07 & 1.40 & 241.74\\
95015 & 201.54907 & -47.87303 & 16.122 & 15.264 & 13.475 & 12.977 & 12.884 & 5076 & 2.56 & 1.28 & 239.10\\
96011 & 201.52316 & -47.88203 & 13.954 & 12.975 & 11.027 & 10.514 & 10.416 & 4894 & 1.56 & 1.53 & 229.55\\
98012 & 201.35549 & -47.89600 & 14.561 & 13.623 & 12.034 & 11.552 & 11.471 & 5210 & 1.96 & 1.43 & 229.93\\
\hline     
\end{tabular}
\end{table*}

\begin{table*}
\caption{Stellar abundances}
\label{t3}
\centering
\begin{tabular}{lccccccccccccc}
\hline
\hline
ID & FeI & ${\rm [FeI/H]}$ & NaI & MgI & SiI & CaI & TiI & TiII & CrI & NiI & ZnI & YII & BaII\\
\hline                      
00006 & 6.15 & -1.35 & 4.91 & 6.27 & 6.63 & 5.25 & 3.94 & 3.89 & 4.11 & 4.61 & 3.35 & 1.06 & 1.27\\
08004 & 6.23 & -1.27 & 5.74 & 6.31 & 6.97 & 5.53 & 4.12 & 4.30 & 4.38 & 5.01 & 3.61 & 1.75 & 1.93\\
10006 & 5.80 & -1.70 &  -   &  -   &  -   & 5.03 & 3.77 & 3.64 & 3.76 &  -   & 3.09 &  -   & 1.57\\
10009 & 6.18 & -1.32 & 5.61 & 6.38 &  -   & 5.27 & 3.93 & 4.03 & 4.24 & 5.04 & 3.04 & 1.26 & 1.69\\
10010 & 6.45 & -1.05 & 5.38 & 6.71 &  -   & 5.65 & 3.91 & 4.02 & 4.65 & 5.07 & 3.42 & 1.86 & 2.08\\
13006 & 5.93 & -1.57 &  -   & 6.31 &  -   & 5.06 & 3.88 & 3.61 & 4.04 &  -   & 2.95 &  -   & 0.37\\
14002 & 6.02 & -1.48 &  -   &  -   &  -   & 5.25 & 3.90 & 3.72 & 4.00 & 5.19 & 3.28 & 0.61 & 1.15\\
22007 & 6.43 & -1.07 & 5.80 & 6.88 & 6.85 & 5.61 & 4.07 & 4.17 & 4.51 & 5.09 & 3.54 & 1.80 & 2.00\\
25004 & 6.14 & -1.36 &  -   &  -   &  -   & 5.44 & 4.09 & 3.81 & 3.95 &  -   &  -   &  -   & 1.08\\
27008 & 6.52 & -0.98 &  -   &  -   & 7.29 & 5.60 & 4.30 & 4.24 & 4.91 &  -   &  -   & 1.86 & 2.71\\
28009 & 6.29 & -1.21 &  -   &  -   &  -   & 5.53 & 4.20 & 4.29 &  -   & 4.97 &  -   &  -   & 0.65\\
33006 & 6.07 & -1.43 &  -   & 6.38 &  -   & 5.30 & 4.10 & 4.27 & 4.32 & 4.73 &  -   &  -   & 1.43\\ 
34008 & 6.11 & -1.39 & 5.52 & 6.24 &  -   & 5.29 & 3.87 & 3.99 & 4.00 & 4.83 &  -   & 1.27 & 1.54\\
38006 & 5.97 & -1.53 &  -   & 6.19 &  -   & 5.15 & 3.79 & 3.88 & 4.16 & 4.99 & 3.19 & 0.65 & 0.61\\
39013 & 6.01 & -1.49 &  -   & 6.51 & 6.85 & 5.30 & 3.77 & 3.71 & 4.13 & 4.80 &  -   & 1.34 & 1.17\\
42012 & 6.10 & -1.40 & 5.05 & 6.62 & 6.66 & 5.42 & 3.93 & 4.08 & 4.22 & 4.85 & 3.37 & 1.62 & 1.61\\
43002 & 5.94 & -1.56 & 5.10 & 6.09 &  -   & 5.24 & 3.96 & 3.81 & 4.29 &  -   &  -   & 1.52 & 1.15\\
45011 & 6.16 & -1.34 & 5.30 & 6.63 & 6.66 & 5.46 & 4.24 & 4.14 & 4.28 & 4.94 &  -   & 0.98 & 1.64\\
45014 & 5.76 & -1.74 &  -   & 6.25 &  -   & 5.00 & 3.64 & 3.65 & 3.83 &  -   &  -   &  -   & 0.10\\
46003 & 5.81 & -1.69 &  -   & 6.04 &  -   & 5.10 & 3.63 & 3.63 & 4.16 & 4.75 & 3.03 & 0.41 & 0.36\\
48009 & 6.24 & -1.26 & 5.30 & 6.83 &  -   & 5.61 &  -   &  -   & 4.58 & 5.18 & 3.68 & 1.10 & 2.06\\
49008 & 6.09 & -1.41 &  -   & 6.43 &  -   & 5.57 & 4.19 & 4.19 & 4.36 & 5.04 & 4.18 & 2.34 & 1.79\\
51005 & 6.08 & -1.42 &  -   & 6.31 &  -   & 5.56 & 4.01 & 4.34 & 4.55 & 4.97 & 3.59 & 1.28 & 1.20\\
57006 & 5.80 & -1.70 & 5.61 &  -   &  -   & 5.10 & 3.97 & 4.11 & 3.84 &  -   & 2.96 & 0.68 & 0.47\\
61009 & 5.76 & -1.74 &  -   & 6.12 & 6.48 & 5.13 & 3.74 & 3.80 & 4.13 & 5.82 &  -   & 0.38 & 0.50\\
76015 & 5.90 & -1.60 &  -   &  -   &  -   & 5.21 & 3.85 & 3.88 & 4.06 &  -   & 3.40 & 1.12 & 1.70\\
77010 & 6.56 & -0.94 & 5.61 & 7.16 & 7.21 & 5.82 & 4.10 & 4.17 & 4.73 & 5.29 & 3.46 & 1.84 & 1.81\\
78008 & 6.14 & -1.36 &  -   &  -   &  -   & 5.15 & 4.10 & 3.92 & 4.32 & 5.06 &  -   & 0.72 & 0.96\\
80017 & 6.04 & -1.46 &  -   &  -   &  -   & 5.23 & 3.64 & 3.79 & 3.87 &  -   &  -   &  -   & 0.55\\
82012 & 5.86 & -1.64 & 5.63 &  -   &  -   & 5.08 & 3.82 & 3.76 & 4.16 &  -   & 3.72 & 1.06 & 1.29\\
85007 & 5.52 & -1.98 &  -   & 6.29 &  -   & 5.14 & 3.67 & 3.30 & 3.83 & 4.74 &  -   & 0.61 & 0.89\\
85014 & 6.03 & -1.47 &  -   &  -   &  -   & 5.50 & 4.24 & 4.05 & 4.42 &  -   &  -   & 1.86 & 1.62\\
85019 & 5.88 & -1.62 &  -   & 6.59 &  -   & 5.29 & 3.96 & 3.99 & 4.19 &  -   & 3.56 & 1.47 & 1.68\\
86007 & 5.84 & -1.66 & 5.72 & 6.24 & 6.87 & 5.50 & 4.05 & 3.86 & 4.43 & 4.80 & 3.67 & 1.24 & 1.65\\
86010 & 5.87 & -1.63 &  -   &  -   &  -   & 5.17 &  -   &  -   &  -   &  -   &  -   &  -   & 0.71\\
86017 & 6.15 & -1.35 & 5.55 &  -   & 6.84 & 5.54 & 4.19 & 4.21 & 4.17 & 5.07 & 3.33 & 1.22 & 2.30\\
87009 & 6.12 & -1.38 &  -   & 6.59 &  -   & 5.62 & 4.41 & 4.11 & 4.83 & 4.90 & 3.41 & 1.98 & 1.84\\
88023 & 6.10 & -1.40 & 5.08 &  -   &  -   & 5.52 & 4.26 & 4.16 & 4.52 & 4.65 &  -   & 1.80 & 2.20\\
89009 & 5.74 & -1.76 &  -   & 6.28 &  -   & 5.05 & 3.55 & 3.76 & 4.09 & 4.55 &  -   & 0.19 & 0.41\\
89014 & 6.14 & -1.36 & 5.84 &  -   & 6.66 & 5.53 & 4.08 & 4.16 & 4.45 & 4.92 &  -   & 1.18 & 1.64\\
90008 & 5.65 & -1.85 & 5.32 & 6.27 &  -   & 5.35 & 3.82 & 3.29 & 4.20 &  -   & 3.53 & 0.73 & 1.00\\
90019 & 5.83 & -1.67 &  -   &  -   &  -   & 5.16 & 4.10 & 3.70 & 4.14 &  -   &  -   & 0.54 & 0.47\\
90020 & 5.89 & -1.61 &  -   &  -   &  -   & 5.09 & 3.70 & 3.72 &  -   &  -   &  -   & 0.57 & 0.60\\
93016 & 6.20 & -1.30 &  -   & 6.69 &  -   & 5.37 & 4.07 & 3.94 & 4.59 &  -   &  -   & 1.85 & 1.49\\
94011 & 5.93 & -1.57 &  -   &  -   &  -   & 5.17 & 4.02 & 4.00 & 4.12 &  -   &  -   & 0.43 & 0.72\\
95015 & 6.24 & -1.26 &  -   & 6.72 &  -   & 5.32 & 4.31 & 4.28 & 4.71 & 5.12 & 3.74 & 1.33 & 1.48\\
96011 & 5.96 & -1.54 & 5.68 &  -   &  -   & 5.25 & 4.20 & 4.05 & 4.33 & 4.59 &  -   & 1.15 & 1.70\\
98012 & 5.85 & -1.65 &  -   &  -   &  -   & 4.98 &  -   &  -   &  -   &  -   &  -   &  -   & 0.68\\
\hline 
Obs. lines & 30 &    & 2    &  1   &  2   &  10  &  10  &  5   &   5  &   5  &  1   &   4  &   2\\      
\hline    
\end{tabular}
\end{table*}

\begin{figure*}
\centering
\includegraphics[width=10cm]{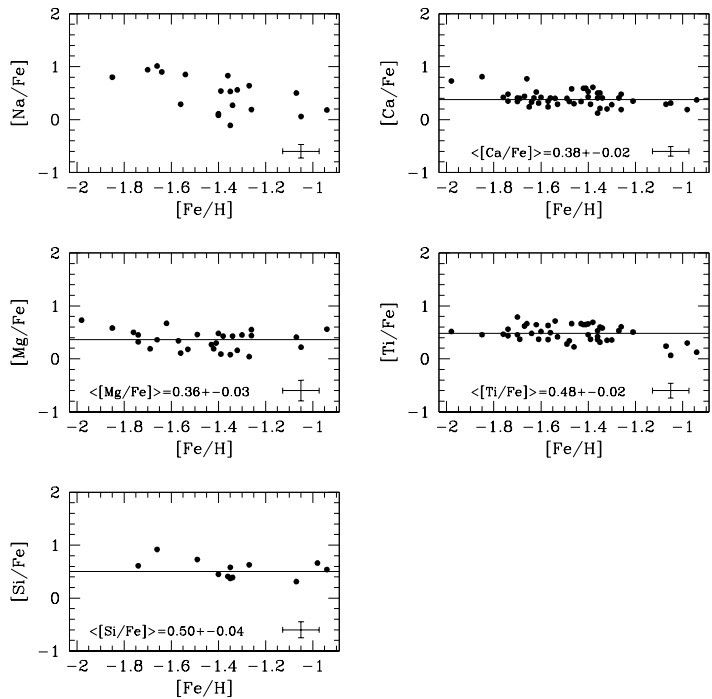}
\caption{Trend of Na and $\alpha-$element abundance ratios as a function of
  [Fe/H]. Mean values (continuous lines) are provided for those elements which do not show a 
  sizable scattering. See also Table~3.}
\label{f3}
\end{figure*}

\begin{figure*}
\centering
\includegraphics[width=10cm]{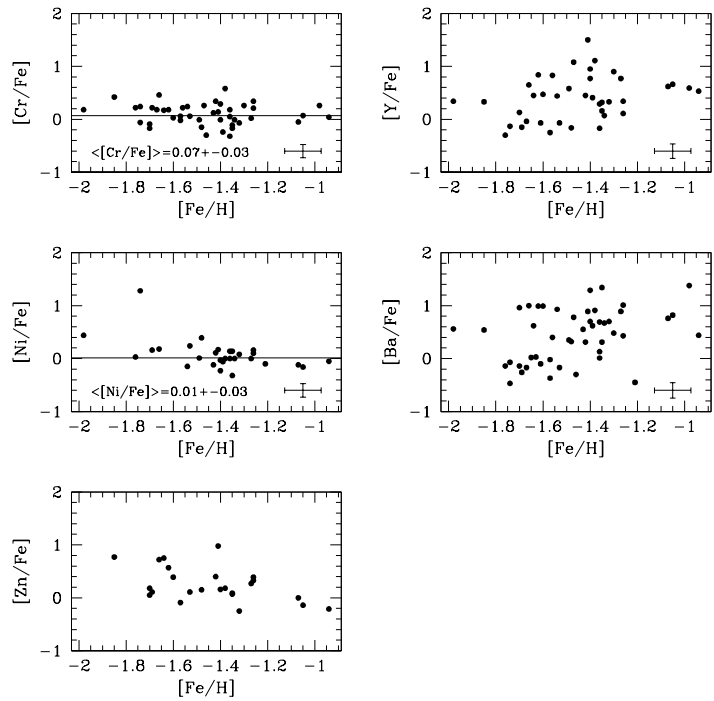}
\caption{Trend of abundance ratios for Iron peak elements (Ni and Cr),
Zn, Y and Ba for the outer region stars.
Mean values (continuous lines) are provided when there is no sizable scatter }
\label{f4}
\end{figure*}

\subsection{The linelist}

The line-lists for the chemical analysis were obtained from the VALD
database (Kupka et al.\ 1999) and calibrated using the Solar-inverse technique.
For this purpose we used the high resolution, high S/N Solar spectrum
obtained at NOAO ($National~Optical~Astronomy~Observatory$, Kurucz et
al.\ 1984). The EWs for the reference Solar spectrum were obtained in the same
way as the observed spectra, with the exception of the strongest lines, where
a Voigt profile integration was used. Lines affected by blends were rejected
from the final line-list.
Metal abundances were determined by the Local Thermodynamic Equilibrium (LTE)
program MOOG (freely distributed by C. Sneden, University of Texas at Austin),
coupled with a solar model atmosphere interpolated from the Kurucz (1992) grids
using the canonical atmospheric parameters for the Sun: $\rm {T_{eff}=5777}$ K,
$\rm {log(g)}=4.44$, $\rm {v_{t}=0.80}$ km/s and $\rm {[Fe/H]=0.00}$.
In the calibration procedure, we adjusted the value of the line strength
log(gf) of each spectral line in order to report the abundances obtained from
all the lines of the same element to the mean value.
The chemical abundances obtained for the Sun and used in the paper as
reference are reported in Tab.~\ref{t1}.

\subsection{Atmospheric parameters}

Estimates of the atmospheric parameters were derived from 
BVJHK photometry. Effective temperatures (T$_{\rm eff}$) for each 
star were derived from the T$_{\rm eff}$-color relations
(Alonso et al.\ 1999, Di Benedetto 1998, and Ramirez \& M\'elendez 2005).  
Colors were de-reddened using a reddening given by
Schlegel et al. (1998). A value E(B-V) = 0.134 mag. was  adopted.

Surface gravities log(g) were obtained from the canonical equation:

\begin{center}
${\rm log(g/g_{\odot}) = log(M/M_{\odot}) + 4\cdot
  log(T_{eff}/T_{\odot}) - log(L/L_{\odot}) }$
\end{center}

For M/M$_{\odot}$ we adopted 0.8 M$_{\odot}$, which is the
typical mass of RGB stars in globular clusters.
The luminosity ${\rm L/L_{\odot}}$ was  obtained from the absolute
magnitude M$_{\rm V}$, assuming an absolute distance modulus 
of (m-M)$_{\rm 0}$=13.75 (Harris 1996), which gives an apparent distance
modulus of (m-M)$_{\rm V}$=14.17 using the adopted reddening. 
The bolometric correction ($\rm {BC}$) was  derived by adopting the relation
BC-T$_{\rm eff}$ from Alonso et al.\ (1999). \\
Finally, microturbolence velocity ($\rm {v_{t}}$) was  obtained from the
relation (Marino et al. 2008):

\begin{center}
${\rm v_{t}\ (km/s) = -0.254\cdot log(g)+1.930}$
\end{center}

which was obtained specifically for old RGB stars, as it is our present sample.
Adopted atmospheric parameters for each star are reported in Tab.~\ref{t2}
in column 9,10,11.
In this Table column 1 gives the ID of the star, columns 2 and 3 the
coordinates, column 4,5,6,7,8 the B,V,J,H,K magnitudes, column 12 the
heliocentric radial velocity.

\subsection{Chemical abundances}

The Local Thermodynamic Equilibrium (LTE) program MOOG (freely
distributed by C. Sneden, University of Texas at Austin) has been used to
determine the abundances from EWs, coupled with model atmosphere
interpolated from the Kurucz (1992) for the parameters obtained as described
in the previous Section.
The wide spectral range of the UVES data allowed us to derive the
chemical abundances of several elements. Chemical abundances
for single stars we obtained are listed in Tab.~\ref{t3}.
The last line of this table gives the mean number of lines we were able to 
measured for each elements.
Ti is the only element for which we could measure neutral and first ionization
lines. The difference of mean abundances obtained from the two stages is:
\begin{center}
${\rm \Delta(TiI-TiII)=0.03\pm0.01}$
\end{center}
This difference is small and compatible with zero within 3 $\sigma$. This confirms
that gravities obtained by the canonical equation are not affected by appreciable
systematic errors.

\subsection{Internal errors associated with the chemical abundances}

The measured abundances of every element vary from
star to star as a consequence of both measurement errors and
intrinsic star to star abundance variations.
In this section our goal is to search for evidence of intrinsic
abundance dispersion in each element by comparing the observed
dispersion $\sigma_{\rm {obs}}$ and that produced by internal errors
($\Delta_{\rm {tot}}$). Clearly, this requires an accurate analysis of
all the internal sources of measurement errors.
We remark here that we are interested in star-to-star intrinsic
abundance variation, i.e. we want to measure the internal intrinsic
abundance spread of our sample of stars. For this reason, we
are not interested in external sources of error which are systematic
and do not affect relative abundances.\\
It must be noted that two main sources of errors contribute
to $\sigma_{\rm {tot}}$. They are:
\begin{itemize}
\item the errors $\sigma_{\rm {EW}}$ due to the uncertainties in the
  EWs measures;
\item the uncertainty $\sigma_{\rm {atm}}$ introduced by errors in the 
  atmospheric parameters adopted to compute the chemical abundances.
\end{itemize}

$\sigma_{\rm {EW}}$ is given by MOOG for each element and each star.
In Tab.~\ref{t4} we reported in the second column the average $\sigma_{\rm {EW}}$
for each element. For Mg and Zn we were able to measure only one line.
For this reason their $\sigma_{\rm {EW}}$ has been obtained as the mean of
$\sigma_{\rm {EW}}$ of Na and Si multiplied by $\sqrt 2$. 
Na and Si lines were selected because their strength was similar to that of Mg
and Zn features. This guarantees that $\sigma_{\rm {EW}}$, due to the photon
noise, is the same for each single line.

Errors in temperature are easy to determine because, for each star, it is
the r.m.s. of the temperatures obtained from the single colours. The mean
error $\Delta$T$_{\rm eff}$ turned out to be 50 K.
Uncertainty on gravity has been obtained by the canonical formula using the
propagation of errors. The variables used in this formula that are affected
by random errors are T$_{\rm eff}$ and the V magnitude. For temperature we
used the error previously obtained, while for V we assumed a error of 0.1 mag,
which is the typical random error for photographic magnitudes. Other error
sources (distance modulus, reddening, bolometric correction) affect
gravity in a systematic way, so are not important to our analysis.
Mean error in gravity turned out to be 0.06 dex. This implies, in turn,  a mean error
in microturbolence of 0.02 km/s.
 
Once the internal errors associated with the atmospheric parameters were
calculated, we re-derived the abundances of two reference stars (\#00006 and
\#42012) which roughly cover the whole temperature range of our sample 
by using the following combination of atmospheric parameters:
\begin{itemize}
\item ($\rm {T_{eff}} \pm \Delta (\rm {T_{eff}})$, $\rm {log(g)}$,  $\rm {v_{t}}$)
\item ($\rm {T_{eff}} $, $\rm {log(g)} \pm \Delta (\rm {log(g)})$,  $\rm {v_{t}}$)
\item ($\rm {T_{eff}} $, $\rm {log(g)}$,  $\rm {v_{t}} \pm \Delta (\rm {v_{t}})$)
\end{itemize}
where $\rm {T_{eff}}$, $\rm {log(g)}$,  $\rm {v_{t}}$ are the measures
determined in Section 3.2 .

The difference of abundance between values obtained with the original
and those ones obtained with the modified values gives
the errors in the chemical abundances due to uncertainties in
each atmospheric parameter. They are listed in Tab.~\ref{t4} (columns 3, 4
and 5) and are the average values obtained from the two stars. 
Because the parameters were not obtained indipendently we cannot
estimate of the total error associated with the abundance
measures by simply taking the squadratic sum of all the single errors.
Instead we calculated the upper limits for the total error as:
\begin{center}
$\rm {\Delta_{tot}=\sigma_{EW}+\Sigma(\sigma_{atm})}$
\end{center}
listed in column 6 of Tab.~\ref{t4}.
Column 7 of Tab.~\ref{t4} is the observed dispersion.
Comparing $\sigma_{\rm obs}$ with $\Delta_{\rm tot}$ (wich is an upper limit) we can see
that for many elements (Mg, Si, Ca, Ti, Cr, Ni) we do not find any evidence of inhomogeneity
among the whole Fe range. Some others (Na, Zn, Y, Ba) instead show
an intrinsic dispersion. This is confirmed also by Figs. \ref{f3} and \ref{f4}
(see next Section).
Finally we just mention here the problem of the differential reddening. 
Some authors (Calamida et al. 2005) claim that is is of the order of 0.03 mag,
while some others (McDonald et al. 2009) suggest a value lower than 0.02 dex.
However all those results concern the internal part, while no information is
available for the region explored in this paper. 
We can only say that an error on the reddening of 0.03 dex would alter
the temperature of 90 degrees.

\begin{table*}
\caption{Internal errors associated with the chemical abundances
due to errors in the EW measurement (column 2) and in the atmospheric
parameters (column 3,4,5) for the studied elements.
6$^{th}$ column gives the total internal error, while the last
column gives the observed dispersion of the abundances. See text for
more details.
}

\label{t4}
\centering
\begin{tabular}{lcccccc}
\hline
\hline
El. & $\sigma_{\rm EW}$ & $\Delta$T$_{\rm eff}$ & $\Delta$log(g) & $\Delta$v$_{\rm t}$ & $\Delta_{\rm tot}$ & $\sigma_{\rm obs}$\\
\hline                      
${\rm [FeI/H]}$    & 0.05 & 0.05 & 0.01 & 0.02 & 0.13 &  -  \\
${\rm [NaI/FeI]}$  & 0.12 & 0.02 & 0.01 & 0.02 & 0.17 & 0.34\\
${\rm [MgI/FeI]}$  & 0.18 & 0.02 & 0.01 & 0.02 & 0.23 & 0.18\\
${\rm [SiI/FeI]}$  & 0.15 & 0.03 & 0.01 & 0.02 & 0.21 & 0.12\\
${\rm [CaI/FeI]}$  & 0.09 & 0.01 & 0.00 & 0.01 & 0.11 & 0.11\\
${\rm [TiI/FeI]}$  & 0.14 & 0.04 & 0.01 & 0.01 & 0.20 & 0.19\\
${\rm [TiII/FeI]}$ & 0.13 & 0.04 & 0.03 & 0.01 & 0.21 & 0.17\\
${\rm [CrI/FeI]}$  & 0.12 & 0.03 & 0.01 & 0.01 & 0.17 & 0.17\\
${\rm [NiI/FeI]}$  & 0.13 & 0.01 & 0.01 & 0.01 & 0.16 & 0.14\\
${\rm [ZnI/FeI]}$  & 0.19 & 0.04 & 0.03 & 0.02 & 0.28 & 0.32\\
${\rm [YII/FeI]}$  & 0.13 & 0.03 & 0.03 & 0.01 & 0.20 & 0.42\\
${\rm [BaII/FeI]}$ & 0.14 & 0.02 & 0.03 & 0.00 & 0.19 & 0.50\\
\hline     
\end{tabular}
\end{table*}

\section{Results of abundance analysis}
The results of the abundance analysis are shown in Fig.~\ref{f2} for [Fe/H],
and in Figs.~\ref{f3} and \ref{f4} for all the abundance ratios we could derive.
A Gaussian fit was used to derive the mean metallicity of the three peaks in
Fig.~\ref{f2}. We found the following values: -1.64 (metal poor
population, {\it MPP}), -1.37 (intermediate metallicity population, {\it IMP}), and -1.02
(metal rich population, {\it MRP}). Stars belonging to each of the three populations are
identified with different colors in Fig.~\ref{f1}.The population mix is in the proportion 
({\it MPP:IMP:MRP}) = (21:23:4).\\

\noindent
The abundance ratio trends versus [Fe/H] 
are shown in the various panels in Figs.~\ref{f3} and \ref{f4} for all the elements we could measure.
When the abundance ratio scatter is low (lower than 0.2 dex which, according to
the previous Section, implies a homogeneous abundance) we also show the mean value
of the data as a continuous line, to make the comparison with literature easier.
What we find in the outer region of $\omega$~ Cen is in basic agreement
with previous investigations. Comparing our trends with -e.g.- Norris \& Da Costa (1995)
values( see next Section for a more general comparison with the literature), 
we find that all the abundance ratios we could measure are in very good
agreement with that study, except for [Ti/Fe], 
which is slightly larger in our stars, and [Ca/Fe], 
which is slightly smaller in our study. However, within the measurement errors we do not
find any significant deviation.\\ 
The $\alpha-$elements (Mg, Ti, Si and Ca, see Fig.~\ref{f3}) 
are systematically overabundant with respect to the Sun, 
while iron peak elements (Ni and Cr, see Fig.~\ref{f4}) are basically solar.
Similarly, overabundant in average with respect to the Sun are Y, Ba and Zn (see Fig.~\ref{f4}). 
Y abundance ratio show some trend with [Fe/H], but of the same sign and 
comparable magnitude to Norris \& Da Costa (1995).

\begin{figure}
\centering
\includegraphics[width=\columnwidth]{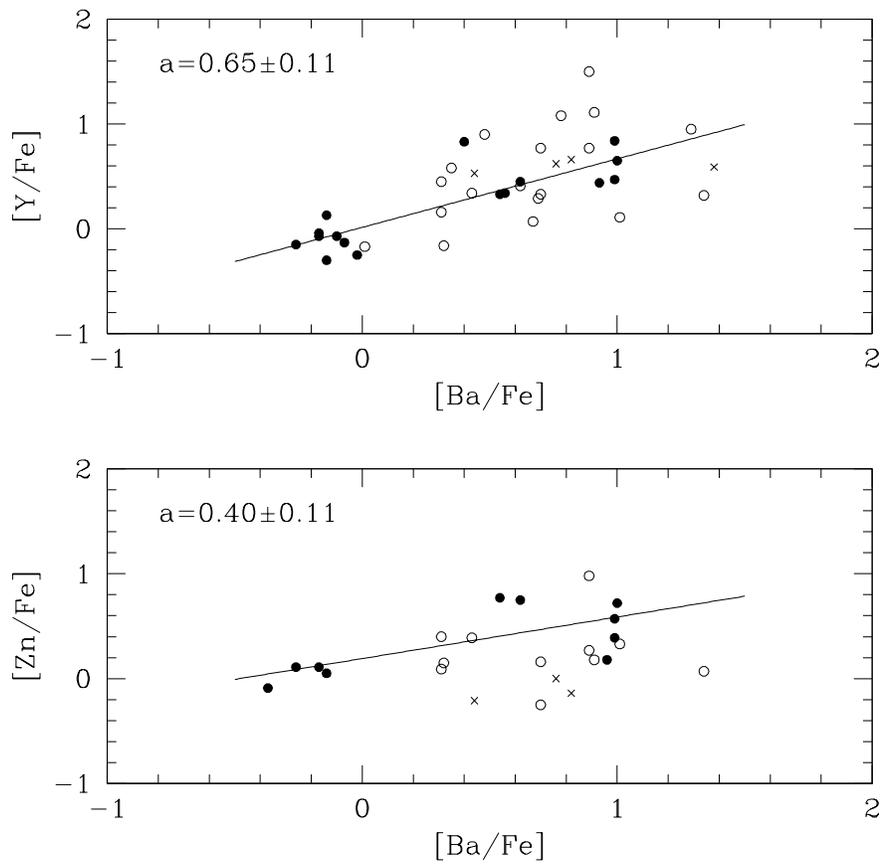}
\caption{
Abundance ratios of [Y/Fe] and [Zn/Fe] vs. [Ba/Fe] for our sample.
Filled circles, open circles, and crosses are MPP, IMP, MRP stars
respectively. A straight line has been fitted to MPP stars. The value of the slope
({\it a}) is given. In both cases {\it a} is out of more than 3$\sigma$ with
respect the null trend value, implying that trends for MPP are real.
}
\label{f5}
\end{figure}

Finally, we looked for possible correlations between abundance ratios, and compare
the outcome from the different populations of our sample. This was possible only for [Y/Fe] and [Zn/Fe]
versus [Ba/Fe], and it is plotted in Fig.~\ref{f5}. For {\it MPP} (filled circles) a trend
appears both for Zn and Y as a function of Ba (see also value of the slope ($a$) in Fig.~\ref{f5}), with
Ba-poor stars being also Zn and Y poor. Y-Ba correlation can be easily
explained because both are neutron-capture elements.\\
As for {\it IMP},  a marginal trend in the Y vs. Ba relation is present,
while no trend appears in the Zn vs. Ba. No trends at all were detected for
MRP, mostly because our sample of {\it MRP} stars is too small for any significant
conclusion. We underline the fact that this different behaviour of {\it MPP} and {\it IMP}
with respect to their Zn-Y-Ba correlations points to a different chemical
enrichment history of the two populations.

\section{Outer versus inner regions}

A promising  application of our data is the comparison of the population mix in the cluster outskirts
with the one routinely found in more central regions of the cluster (Norris \& Da Costa 1995; Smith et al. 2000;
Villanova et al 2007; Johnson et al. 2009; Wylie-de Boer et al. 2009).\\

To this aim, we firstly compute the fraction of stars
in the various metallicity  ([Fe/H]) populations, and compare it with the inner
regions trends from Villanova et al. (2007), for the sake of homogeneity,
to statistically test the significance of their similarity or difference. 
We are aware that this is not much more than a mere exercise.
Firstly, while our program stars are mostly in the RGB phase, in Villanova et al (2007)
sample only SGB stars are present. This implies that we are comparing stars in
slightly different evolutionary phases.
Second, and more important, the statistics is  probably too poor. 
In fact, we report in Table~5 (column 2 and 3) the number of stars
in the different metallicity bin derived from a Gaussian fit to our and Villanova et al. (2007)
data. They  have large errors. We see that within these errors the population mix is basically the
same in the inner and outer regions. Therefore, with so few stars we cannot
detect easily differences between the inner and outer regions.
To check for that, we make use of the Kolmogorov-Smirnov statistics,
and compare the metallicity distributions of the inner and outer samples, to see
whether they come from the same parental distribution. We found that the probability
that the two distributions derive from the same underlying distribution is 77$\%$.
This is quite a small number, and simply means that with these samples we cannot
either disprove or confirm the null hypothesis (say that the two populations have
same parental distribution).
Besides, our sample and that of Villanova et al (2007) do not have stars
belonging to the most metal-rich population of Omega centauri (at
[Fe/H]$\sim$-0.6), which therefore we cannot comment on.\\

\noindent
That clarified, we then compare in Fig.~6 and Fig.~7 the trend of the various elements we could
measure (see Table~4) in the cluster outskirts with the trends found in the central regions by other
studies. In details, in all Fig.~6 panels we indicate with filled circles the data
presented in this study and  with open circles data from Villanova et al. (2007). Crosses indicate
Wylie-de Boer et al. (2009), stars Norris \& Da Costa (1995), empty squares 
Smith et al. (2000) and, finally, empty pentagons Johnson et al. (2009).
We separate in Fig.~6 elements which do no show significant scatter (see Table~4) from
elements which do  show a sizeable scatter (see Fig.~7).
Ba abundances from Villanova et al. (2007) were corrected of $\sim$-0.3 dex,
to take into account the hyperfine structure that seriously affects the Ba
line at 4554 \AA.

\begin{table}
\centering
\label{t5}
\begin{tabular}{ccc}
\hline
Population & Inner & Outer \\
 & $\%$ & $\%$ \\
\hline
MPP & 46$\pm$10 & 45$\pm$10 \\
IMP & 36$\pm$10 & 47$\pm$10 \\
MRP & 18$\pm$10 &  8$\pm$10 \\
\hline
\end{tabular}
 \caption{Percentages of different metallicity populations in the inner and outer regions
of $\omega$~ Cen.}
 \end{table}

\noindent
Looking at Fig.~6, we immediately recognize two important facts.\\
First, all the studies we culled from the literature for Omega Cen central regions
show the same trends.\\
Second, and more important for the purpose of this paper,
we do not see any significant difference bewteen the outer (filled circles)
and inner  (all the other symbols) regions of the cluster. Only Ti seems to be
slightly over-abundant in the outer regions with respect to the more central ones.

\begin{figure}
\centering
\includegraphics[width=\columnwidth]{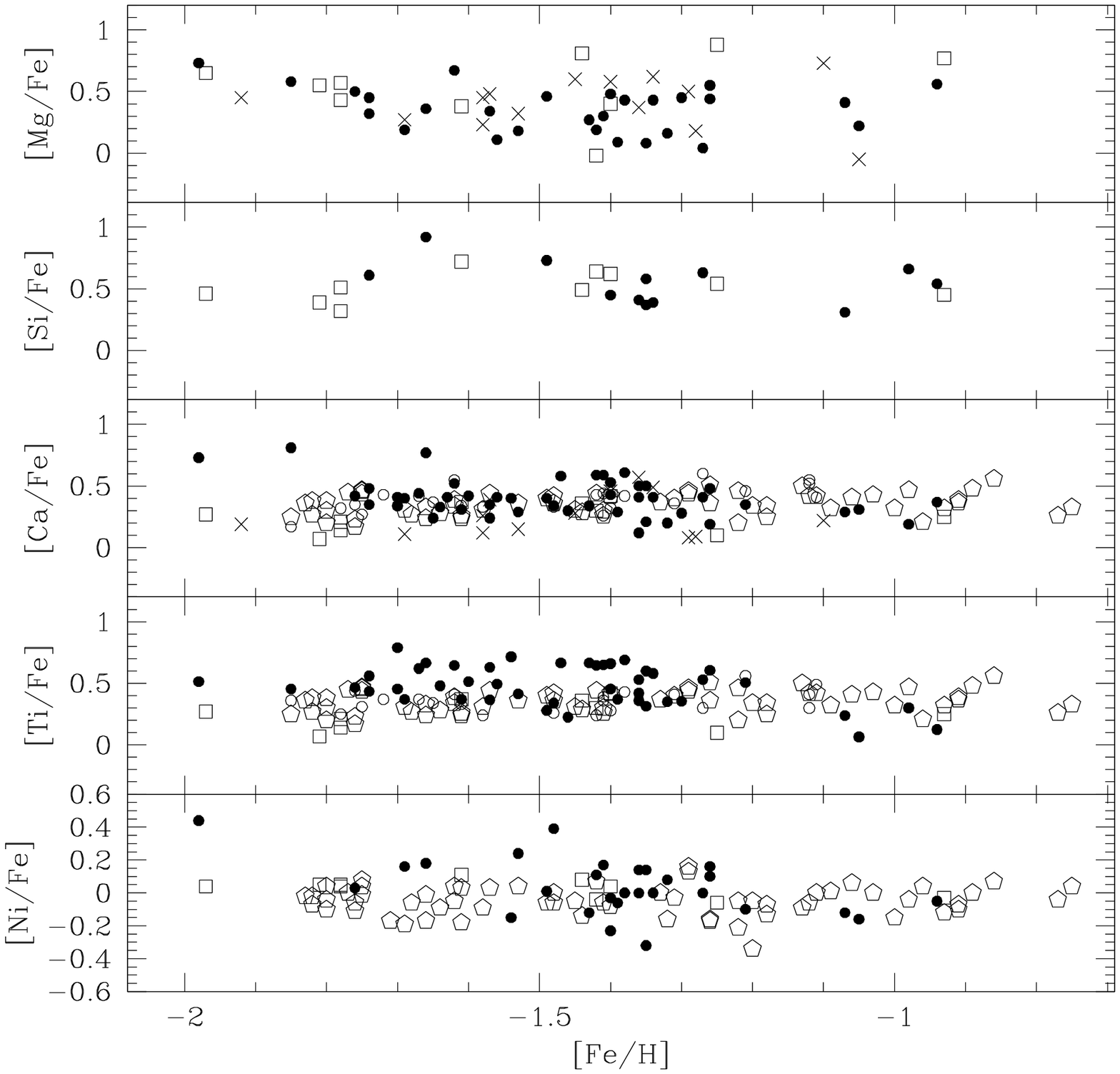}
\caption{Comparison of abundance ratios in the inner and
outer stars (filled circles). Symbols are as follows. Empty circles (Villanova et al. 2007),  crosses 
(Wylie-de Boer et al. 2009), stars (Norris \& Da Costa 1995), empty squares 
(Smith et al. 2000) empty pentagons (Johnson et al. 2009)}
\label{f6}
\end{figure}

\begin{figure}
\centering
\includegraphics[width=\columnwidth]{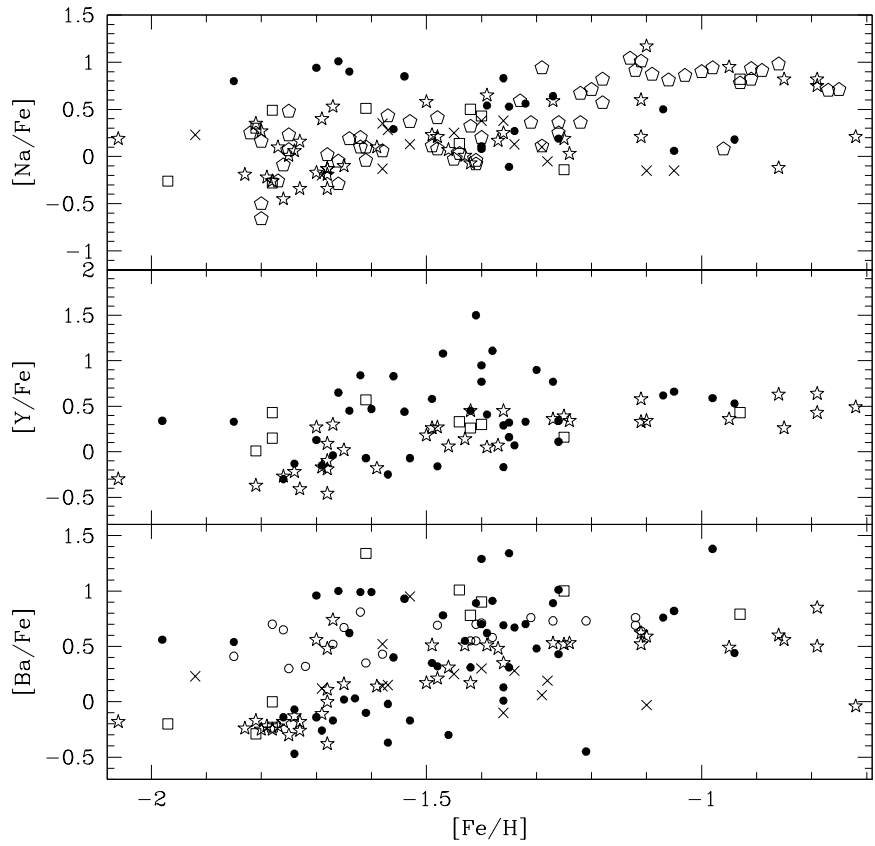}
\caption{Comparison of abundance ratios in the inner
and outer stars (filled circles). Symbols are as in Fig.~6}
\label{f7}
\end{figure}

\begin{figure}
\centering
\includegraphics[width=\columnwidth]{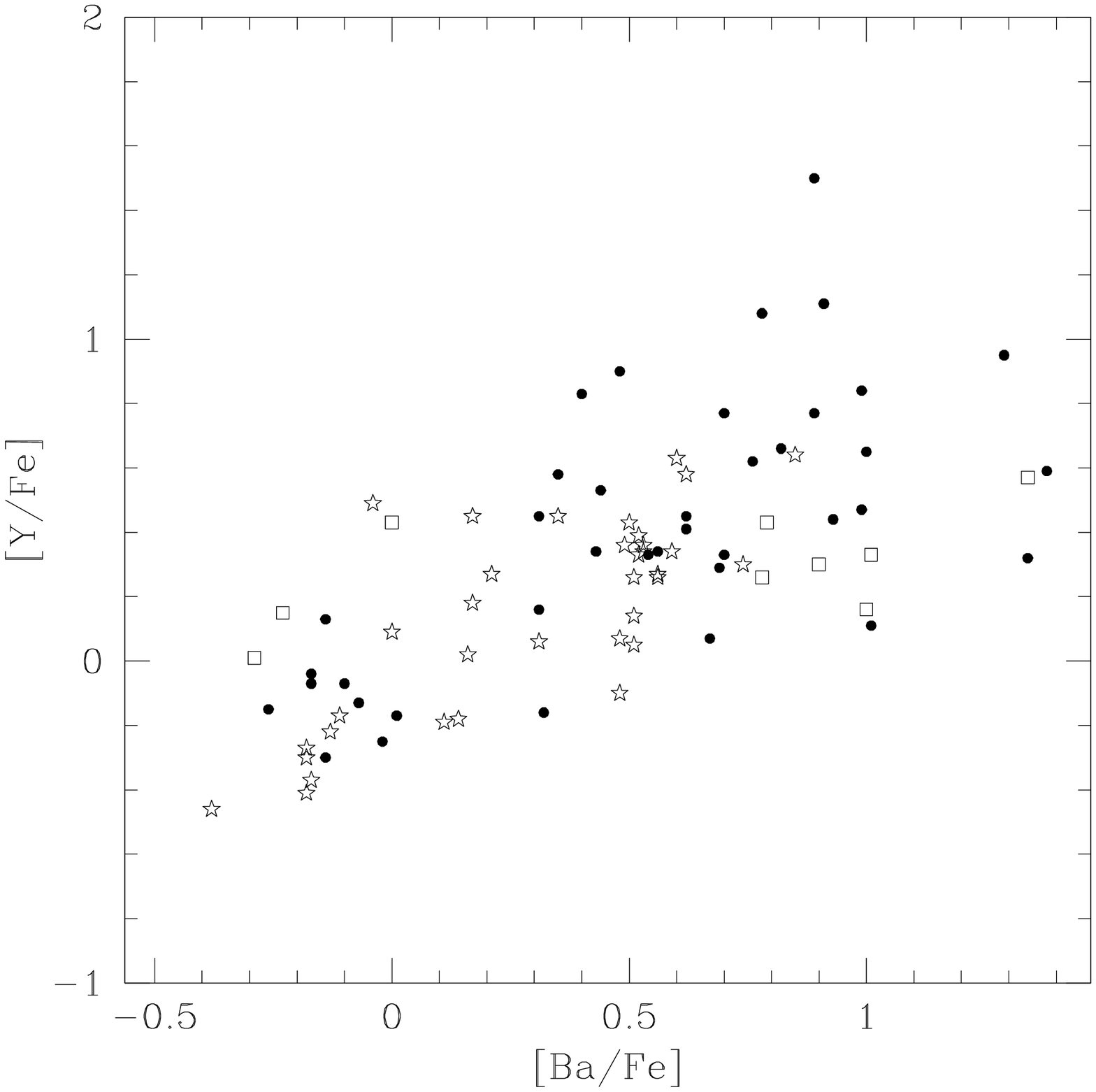}
\caption{Comparison of our Y and Ba abundance ratios (our sample, filled
  circles) with the literature (inner sample).Symbols are as in Fig.~6}
\label{f8}
\end{figure}

\noindent
As for the more scattered elements (see Fig.~7) we notice that Na shows  the opposite trend
in the outer regions with respect to the inner ones, but this is possibly related
to a bias induced by the signal to noise of our data which does not allow us to detect
Na-poor stars in the metal poor population.
On the other hand, Y and Ba do not show any spatial difference. \\

\noindent
Interestingly enough, at low metallicity Ba shows quite a significant scattered
distribution, expecially for stars more metal-poor than -1.2 dex, covering a
range of 1.5 dex. This behaviour is shared with Y and Na, althought at a lower level.
Furthermore, looking carefully at Fig. 4, it is possible to see a hint of
bimodality for the Ba content of the stars having [Fe/H]$<$-1.5 dex
(i.e. belonging to the MMP), with the presence of a group of objects having
[Ba/Fe]$\sim$1.0 dex, and another having [Ba/Fe]$\sim$-0.2 dex.
The same trend is visible in all the literature data.\\
We remind the reader that such a bimodal distribution is similar to that found by 
Johnson et al. (2009, thier Fig.~8) for Al.\\

\noindent
Finally, we compare our Y vs. Ba trend with literature in Fig. 8. Also in this
case the agreement is very good and no radial trend is found.\\

\noindent
The stars studied by Wylie-de Boer et
al. (2009) deserve special attention. They belong to the Kapteyn Group, 
but their kinematics and chemistry suggest a likely association with $\omega$ Cen.
These stars, in spite of being part of a moving group, exhibit 
quite a large iron abundance spread (see Fig. 6 and 7), totally compatible with the one of
$\omega$ Cen. Also their Na and Ba abundance (see Fig. 7) are comparable with
those of the cluster. We suggest that the comparison with our data reinforces
the idea that the Kapteyn Group is likely formed by stars stripped away from $\omega$ Cen.

\section{Conclusions}
In this study, we analized a sample of 48 RGB stars located half-way to the tidal
radius of $\omega$ Cen, well beyond any previous study devoted to the detailed chemical composition of the
different cluster sub-populations.\\
We compared the abundance trends in the cluster outer regions with literature studies which focus
on the inner regions of  $\omega$ Cen.\\

\noindent
The results of this study can be summarized as follows:

\begin{description}
\item $\bullet$ we could not highlight any difference between the outer and inner regions
as far as [Fe/H]is concerned: the same mix of different iron abundance population is present
in both locations;
\item $\bullet$ most elements appear in the same proportion both in the inner and in the outer
zone, irrespective of the particular investigation one takes into account for the comparison;
\item $\bullet$ [Ba/Fe] shows an indication of a bimodal distribution at low metallicity at any location
in the cluster, which deserves further investigation;
\item $\bullet$ no indications emerge of gradients in the radial abundance trend of the elements we could
measure.
\end{description}

\noindent
Our results clearly depend on a small data-set, and more extended studies are encouraged to confirm
or deny our findings.

\section*{Acknowledgements}
Sandro Villanova acknowledges ESO for financial support during several visits to the Vitacura
Science office. The authors express their gratitude to Kenneth Janes for reading carefully
the manuscript.


\end{document}